\documentclass{article}
\title{Dimension Compactification -- a Possible 
Explanation for Superclusters and for Empirical Evidence Usually
Interpreted as Dark Matter} 

\author{Vladik Kreinovich\\
NASA PACES Center\\
University of Texas at El Paso\\
El Paso, TX 79968, USA\\
vladik@cs.utep.edu}

\date{}

\begin{document}
\maketitle

\begin{abstract}
According to modern quantum physics, at the microlevel, the dimension
of space-time is $\ge 11$; we only observe 4 dimensions because the
others are compactified: the size along each of the other dimensions
is much smaller than the macroscale. 
There is no universally accepted explanation of why exactly 4 dimensions
remain at the microscopic level. A natural suggestion is: maybe there
is no fundamental reason why exactly 4 dimensions should remain, maybe 
when we go to even larger scales, some of
the 4 dimensions will be compactified as well? 
In this paper, we explore the consequences of the compactification 
suggestion, and we 
show that -- on the qualitative level -- these consequences have actually
been already observed: as superclusters and as evidence that is
usually interpreted as justifying dark matter. 
Thus, we get a new
possible explanation of both superclusters and dark matter evidence -- 
via dimension compactification. 
\end{abstract}

\medskip

\noindent{\bf Main idea.} 
According to modern quantum physics, at the microlevel, 
space-time has at least 11 dimensions; we only observe 4 
dimensions in our macroobservations because the rest are {\it
compactified:} the size along each of the remaining directions is so
small that on macrolevel, these dimensions can be safely ignored; see, 
e.g., \cite{Green 1988,Polchinski 1998}. 

There is no universally accepted explanation of why exactly 4 dimensions
remain at the microscopic level. A natural suggestion is: maybe there
is no fundamental reason why exactly 4 dimensions should remain. Maybe 
when we go to even larger scales, some of
the 4 dimensions will be compactified as well? 

Could one rigorously argue for ``compactification" when these effects must
occur over the accessible, kiloparsec and larger, scales? In modern
physics, indeed, compactification is related to quantum-size
distances, but there is nothing inherently quantum in the
compactification idea. Indeed, the very first paper that proposed
compactification -- the 1938 paper by A. Einstein and P.~Bergmann
\cite{Einstein 1938} -- described it as simply the extra dimension
being a circle, so that the entire space-time looks like a thin
cylinder whose width is negligible if we operate at large enough
scales. 

In this paper, we explore the consequences of the compactification 
suggestion, and we 
show that -- on the qualitative level -- these consequences have actually
been already observed: as superclusters and as evidence that is
usually interpreted as justifying dark matter. Thus, we get a new
possible explanation of both superclusters and dark matter evidence -- 
via dimension compactification. 
\medskip

\noindent{\bf Geometric consequences of the main idea.} If our idea is 
correct, then, as we increase the scale, we will observe how a 3D picture 
is replaced by a 2D and then by a 1D one. This is exactly what we
observe: while at a macrolevel, we see a uniform distribution of
galaxies in a 3D space, at a larger-scale level, galaxies start forming
{\it superclusters} -- long and thin strands of clusters and
galaxies; see, e.g., \cite{Fairall 1998}. Superclusters are either 
close to a 2D shape (as the ``Great Wall" discovered in the
1980s) or close to 1D. 
\medskip

\noindent{\bf Towards physical consequences of the main idea.} 
How does the
change in dimension affect physics? In non-relativistic
approximation, the gravitation potential $\varphi$ 
is related to the mass density $\rho$ by the Laplace equation 
$\triangle\varphi=\rho$. In the 3D space, this leads to Newton's potential
$\sim 1/r$, with the force 
$$F(r)={G_0\cdot m_1\cdot m_2\over r^2};$$ 
in a 2D space, we get potential $\sim\log(r)$, 
with the force $$F(r)={G_1\cdot m_1\cdot m_2\over r}$$ for some constant
$G_1$. For intermediate scales, it is reasonable to consider a
combination of these two forces:
$$F(r)={G_0\cdot m_1\cdot m_2\over r^2}+{G_1\cdot m_1\cdot m_2\over
r}.\eqno{(1)}$$
In the Appendix, we explain, on the qualitative level, why such combination
naturally follows from the above compactification idea. 

Let us consider the simplest possible compactification (along the
lines of the original Einstein-Bergmann paper), where the space-time
is wrapped as a cylinder of circumference $R$ 
along one of the coordinates. What will the
gravitational potential look like in this simple model? The easiest
way to solve the corresponding Newton's equation in this cylinder is
to ``unwrap" the cylinder into a full space. After this ``unwrapping",
each particle in a cylindrical space (in particular, each source of
gravitation) is represented as infinitely many different bodies at
distances $R$, $2R$, etc., from each other. For further simplicity,
let us consider the potential force between the two bodies on the
wrapping line at 
a distance $r$ from each other. The effect of the second body
on the first one in cylindrical space is equivalent to the joint effect 
of multiple copies of the second body in the unwrapped space:

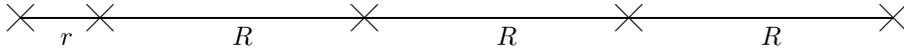
\begin{center}
\begin{picture}(330,50)(0,0)
\put(0,20){\line(1,0){30}}
\put(30,20){\line(1,0){100}}
\put(130,20){\line(1,0){100}}
\put(230,20){\line(1,0){100}}
\put(-5,15){\line(1,1){10}}
\put(-5,25){\line(1,-1){10}}
\put(25,15){\line(1,1){10}}
\put(25,25){\line(1,-1){10}}
\put(125,15){\line(1,1){10}}
\put(125,25){\line(1,-1){10}}
\put(225,15){\line(1,1){10}}
\put(225,25){\line(1,-1){10}}
\put(325,15){\line(1,1){10}}
\put(325,25){\line(1,-1){10}}
\put(15,10){$r$}
\put(80,10){$R$}
\put(180,10){$R$}
\put(280,10){$R$}
\end{picture}
\end{center}

The resulting gravitational potential of a unit mass 
can be described as a sum of
potentials corresponding to all these copies, i.e., 
$$\varphi(r)={1\over r}+{1\over r+R}+{1\over r+2R}+\ldots+{1\over 
r+k\cdot R}+\ldots\eqno{(2)}$$
From the purely mathematical viewpoint, 
this sum is infinite. From the physical viewpoint, however, 
the actual potential is not
infinite: due to relativistic effects, at the current moment of time
$t_0$, the influence of a 
source at a distance $d=r+k\cdot R$ is determined by this source's
location at a time $t_0-d/c$ (where $c$ is the speed of light). Thus,
we only need to add the terms for which $d/c$ is smaller than the age
of the Universe. As a result, we can ignore the slowly increasing
infiniteness of the sum when $k\to\infty$.

How can we estimate this potential? The formula (2) has the form 
$$f(0)+f(1)+\ldots+f(k)+\ldots,$$
where we denoted $f(x)\stackrel{\rm def}{=}
1/(r+x\cdot R)$. It is difficult to get an
analytical expression for the exact sum, but we can use the fact that
this sum is an integral sum for an integral
$\displaystyle\int_0^\infty f(x)\,dx$; this
integral has an analytical expression -- it is 
${\rm const}-(1/R)\cdot \ln(r)$. In 
this approximation
$$\int_0^\infty f(x)\, dx\approx f(0)+f(1)+\ldots+f(k)+\ldots,\eqno{(3)}$$
we used the fact that 
$$\int_0^\infty f(x)\, dx=\int_0^1 f(x)\, dx+\int_1^2 f(x)\,
dx+\ldots+\int_k^{k+1} f(x)\, dx+\ldots,$$
and we approximated each term $\displaystyle\int_k^{k+1} f(x)\, dx$ by
$f(k)$. This approximation is equivalent to approximating the function 
$f(x)$ on the interval $[k,k+1]$ by its value $f(k)$ at the left
endpoint of this interval -- i.e., by the first term in the Taylor
expansion of the function $f(x)$ around the point $k$. A natural next
approximation is when instead of only taking 
the first term, we consider the first {\it two} terms in this Taylor
expansion, i.e., when on each interval $[k,k+1]$, 
we approximate the function $f(x)$ by a formula $f(k)+f'(k)\cdot
(x-k)$. Under this approximation, 
$$\int_k^{k+1} f(x)\, dx \approx f(k)+{1\over 2}\cdot f'(k),$$
and therefore, 
$$\int_0^\infty f(x)\, dx=
\int_0^1 f(x)\, dx+\int_1^2 f(x)\,
dx+\ldots+\int_k^{k+1} f(x)\, dx+\ldots=$$
$$(f(0)+f(1)+\ldots+f(k)+\ldots)+{1\over 2}\cdot
(f'(0)+f'(1)+\ldots+f'(k)+\ldots).\eqno{(4)}$$
The second term in the right-hand side can be (similarly to the
formula (3)) 
approximated by 
$$f'(0)+f'(1)+\ldots+f'(k)+\ldots\approx {\rm const}+\int_0^\infty
f'(x)\,dx-{1\over 2}\cdot f(0).$$ 
Thus, from the formula (4), we can conclude that 
$$f(0)+f(1)+\ldots+f(k)+\ldots\approx 
\int_0^\infty f(x)\, dx+f(0).$$
In particular, for our function $f(x)$, we get 
$$\varphi(r)=f(0)+f(1)+\ldots+f(k)+\ldots\approx {\rm const}-
{1\over R}\cdot
\ln(r)+{1\over 2}\cdot {1\over r}.$$
Differentiating relative to $r$, we get the desired formula (1) for the
gravitational force: 
$$F(r)=-{d\varphi(r)\over dr}={1\over R\cdot r}+{1\over 2r^2},$$
with $R=2G_0/G_1$. According to estimates from \cite{Kuhn 1987}, we
expect $R$ to be between $\approx 10$ and $\approx 30$ kpc. 
\medskip

\noindent{\bf Physical consequences of the main idea.} 
The force described by formula (1) 
is exactly the force that, according to Kuhn, Milgrom, et al. 
\cite{Kuhn 1987,Milgrom 1983,Milgrom 2002,Sanders 2002}, 
is empirically needed to describe the observations if we want to avoid 
dark matter. Indeed, in Newtonian mechanics, for any large-scale rotating 
gravitational system, if we know the rotation speed $v$ at a distance
$r$ from the center, we can find the mass $M_g(r)$ inside the sphere of
radius $r$ by equating the acceleration $v^2/r$ with the
acceleration $G_0\cdot M_g(r)/r^2$ provided by the Newton's law. As a
result, we get $M_g(r)=r\cdot v^2/G_0$. Alternatively, we can also count
masses of different observed bodies and get $M_L(r)$ -- the total mass of 
luminescent bodies. It turns out that $M_L(r)\ll M_g(r)$ -- hence the
traditional explanation that in addition to luminescent bodies, there 
is also ``dark" (non-luminescent) matter. 

An alternative explanation is not to introduce any new unknown type
of matter -- i.e., assume that $M_g(r)\approx M_L(r)$ --
but rather 
change the expression for the force, 
or, equivalently, assume that the gravitational constant 
$G_0$ is not a constant but it may
depend on $r$: $G_0=G(r)$. Equating 
the acceleration $v^2/r$ with the
acceleration $G(r)\cdot M_L(r)/r^2$ provided by the new gravity law, 
we can determine $G(r)$ as $G(r)=v^2\cdot r/M_L(r)$. Observation data
show that $G(r)=G_0+G_1\cdot r$ for some constant $G_1$ 
-- i.e., that the dependence of the gravity force on distance 
is described by the formula 
$$F(r)={G(r)\cdot m_1\cdot m_2\over r^2}=
{G_0\cdot m_1\cdot m_2\over r^2}+{G_1\cdot m_1\cdot m_2\over
r},$$
which is exactly what we deduced from our dimension compactification
idea. 

This idea has been proposed 20 years ago, and one of the reasons why
it has not been universally accepted is that it was difficult to get a
natural 
field theory explanation of this empirical law. We have just shown
that such an empirical explanation comes naturally if we consider the
possibility of dimension compactification. 

This explanation is in line with Milgrom's own explanation; see, e.g., 
\cite{Milgrom 2003}. 

As we go further up in scale, one more dimension starts compactifying,
so we start getting a 1D space in which Laplace equation leads to potential 
$\varphi(r)\sim r$. Thus, at a large-scale level, we should have a
term proportional to $r$ added to the normal gravity potential
formulas. This additional term is exactly what is add when we take a 
cosmological constant $\Lambda$ into consideration.
\medskip

\noindent{\bf Observable predictions of our new idea.} 
A possible observable 
consequence of the additional term $\varphi(r)\sim r$ 
is that it leads to an additional constant term in the gravitational
force and therefore, to a formula 
$G(r)=G_0+G_1\cdot r+G_2\cdot r^2$. Thus, if the empirical dependence
of $G(r)$ on $r$ turns out to be not exactly linear but rather
slightly quadratic, it will be a strong argument in favor of our
compactification idea. 
\medskip

\noindent{\bf Natural open questions.} In this paper, we simply
formulate the idea and explain why we believe this idea to be
prospective. Many related questions are still open:
\begin{itemize}
\item how to concoct a geometry that accomplishes ``compactification" --
what would a metric look like that transitions from large to small
scales? 
\item could we solve for the force law in such a geometry with some
rigor?  
\item will such a profound geometrical
perturbation as the one we propose 
have other consequences beyond a force law
change? perhaps not, but it is worth investigating.
\end{itemize}
\medskip

\noindent{\bf Acknowledgments.}
This work was supported in part by NASA under cooperative 
agreement NCC5-209, by the 
Future Aerospace Science and Technology Program (FAST) 
Center for Structural Integrity of Aerospace Systems,
effort sponsored by the Air Force Office of Scientific Research, Air Force
Materiel Command, USAF, under grant
F49620-00-1-0365, 
and by NSF grants EAR-0112968 and EAR-0225670.

The author is very thankful to Jeffrey R. Kuhn for his ideas and for 
his valuable advise.

\end{document}